\documentclass[showpacs,prb,floatfix,twocolumn,amsmath,a4]{revtex4}

\usepackage{graphicx}

\usepackage{amssymb,amsfonts,amsmath}

\begin{document}

\title{Evidence for the formation of a Mott state in potassium-intercalated pentacene}

\author{M.~F. Craciun$^{1,5}$, G. Giovannetti$^{2,3}$, S. Rogge$^1$,  G. Brocks$^3$,  A.~F. Morpurgo$^{1,6}$ and J. van den Brink$^{2,4}$}

\affiliation{$^1$ Kavli Institute of Nanoscience, Delft University
of Technology, Lorentzweg~1, 2628\,CJ Delft, The Netherlands\\
$^2$Institute Lorentz for Theoretical Physics, Leiden University, Leiden, The Netherlands\\
$^3$Faculty of Science and Technology and MESA+ Institute for Nanotechnology, University of Twente, Enschede, The Netherlands\\
$^4$Institute for Molecules and Materials, Radboud University,
Nijmegen, The Netherlands\\
$^5$Department of Applied Physics, The University of Tokyo, Tokyo
113-8656, Japan\\
$^6$DPMC and GAP, University of Geneva, quai Ernest-Ansermet 24,
CH-1211 Geneva 4, Switzerland}

%%%%%%%%%%%%%%%%%%%%%%%%%%%%%%%%%%%%%%%%%%%%%%%%%%%%%%%%%%%%%%%%

\begin{abstract} We investigate electronic transport through pentacene thin-films intercalated with potassium.
From temperature-dependent conductivity measurements we find that
potassium-intercalated pentacene shows metallic behavior in a
broad range of potassium concentrations. Surprisingly, the
conductivity exhibits a re-entrance into an insulating state when
the potassium concentration is increased past one atom per
molecule. We analyze our observations theoretically by means of
electronic structure calculations, and we conclude that the
phenomenon originates from a Mott metal-insulator transition,
driven by electron-electron interactions.
\end{abstract}

\date{\today}


%\keywords{plastic electronics |thin-films | pentacene| Mott insulator| alkali doping}

\maketitle

\section{Introduction}

Pentacene (PEN) is a conjugated molecule very well-known in the
field of plastic electronics for its use in high-mobility organic
thin-films transistors~\cite{Nelson98,Dimitrakopoulos02}. Plastic
electronics applications rely on the fact that at low density of
charge carriers pentacene films effectively behave as weakly doped
semiconductors~\cite{Sirringhaus00,Gelinck04,Klauk07}. In this
regime, which is studied extensively, the interactions between
charge carriers can be neglected. However, the opposite regime of
high carrier density has remained virtually unexplored. Since
pentacene froms a molecular solid with narrow bandwidth it can be
expected that at high density the simple assumptions of
independent electron-band theories break down and the electronic
correlations determine the electronic properties of the
material~\cite{Brocks04}, as it happens in other molecular
systems. The origin of correlated behavior in these systems is the
competition between the energy gained by delocalizing the $\pi$
electrons (given by the electronic bandwidth) and the Coulomb
repulsion between two carriers on the same molecule. If the
repulsion energy is larger than the one gained on delocalization,
then the electrons become localized and a Mott metal--insulator
transition takes place. Among the most studied molecular systems
in the high carrier density regime are the intercalated C$_{60}$
crystals~\cite{Takenobu00} and the organic charge transfer salts,
where the electronic interactions lead to the appearance of highly
correlated magnetic ground states and unconventional
superconductivity~\cite{Ishiguro,Imada98}.

Our goal is to investigate pentacene compounds at high carrier
density --of the order of one carrier per molecule--  and to show
that electron correlation effects are crucial to understand the
resulting electronic properties. To this end, we have studied
electronic transport through high-quality pentacene thin-films
similar to those used for the fabrication of field-effect
transistors. In order to reach high densities of charge carriers
we intercalate the pentacene films with potassium atoms, to form
K$_x$PEN. Several past experimental studies have addressed the
possibility to chemically dope pentacene thin films through the
inclusion of alkali atoms and iodine. For all these compounds the
structural investigations have shown that large concentrations of
atoms (up to three iodine atoms per pentacene molecule) can
intercalate in between the planes of the pentacene molecular
films. Similarly to the case of intercalated
$C_{60}$~\cite{Forro01}, the alkali atoms donate their electrons
to the Lowest Unoccupied Molecular Orbital (LUMO) of the pentacene
molecules, whereas the iodine donate holes to the Highest Occupied
Molecular Orbital (HOMO)), enabling the control of the
conductivity of the films. 
Earlier studies~\cite{Minakata91,Minakata92,Ito04,Minakata93,Matsuo04a,Matsuo04b,Kanedo05,Fang05} indicate that upon 
iodine or rubidium intercalation the conductivity of pentacene films can 
become large (in the order of 100 $S/cm$) and exhibit a metallic temperature dependence.
The experiments so far, however, have not led to an understanding of the doping
dependence of the conductivity (e.g. how many electrons can be
transferred upon intercalation) or about the microscopic nature of
electrical conduction of doped pentacene in the high carrier
density regime.

In this paper we present the first experimental investigation of
the evolution of the temperature-dependent conductivity of
potassium-intercalated pentacene thin films, with increasing their
doping concentration. We find that, upon K intercalation, PEN
films become metallic in a broad range of doping concentrations,
up to K$_1$PEN, after which the conductivity re-enters an
insulating state. Our experiments also show that the structural
disorder of PEN films plays an important role on the transport
properties of K$_x$PEN, as films of poor structural quality do
not exhibit metallic behavior. The analysis of our data shows that
at high carrier density the conductivity of K$_1$PEN cannot be
described in terms of independent electrons filling the molecular
band originating from the LUMO. Rather, our observations are
consistent with the formation of a Mott insulating state, driven
by electron-electron interactions, as we show theoretically by
calculating the electronic structure of K$_1$PEN.

\section{Preparation of K$_x$PEN Films}

Our choice of working with pentacene thin-films (as opposed to
single crystals) is motivated by both the relevance for
applications --control of doping in organic semiconductors is
important for future plastic electronic devices-- and by the
difficulty to grow crystals of alkali-intercalated pentacene,
which has so far impeded this sort of investigations. All
technical steps of our experimental investigations including the
deposition of pentacene films, potassium intercalation, and
temperature-dependent transport measurements have been carried out
in ultra-high vacuum (UHV) ($10^{-11}$ mbar) in a fashion that is
similar to our previous studies of intercalated phthalocyanine
films~\cite{Craciun06}. The use of UHV prevents the occurrence of
degradation of the doped films over a period of days.

As with the phthalocyanines, the PEN films ($\sim$ 25nm thick)
were thermally evaporated from a Knudsen cell onto a silicon
substrate kept at room temperature. In order to minimize the
parallel conduction through the silicon, we use a high resistive
silicon-on-insulator (SOI) wafer as substrate. The SOI wafer
consists of 2$\mu$m silicon top-layer electrically insulated by
1$\mu$m-thick SiO$_{2}$ layer from the silicon
substrate~\cite{Craciun06}. Ti/Au electrodes (10nm Ti and 50nm Au)
were deposited {\it ex situ} on the SOI substrates, see
Fig.~\ref{fig:fig1}b. After the deposition of the electrodes and
prior to loading the substrate into the UHV system, a
hydrogen-terminated Si surface was prepared by dipping the SOI
surface in a hydrofluoric acid solution and rinsing in de-ionized
water. The use of such a H-terminated silicon surface proved
necessary to achieve sufficient quality in film morphology, as we
will discuss in the next section in more detail.

Special care was taken to chemically purify pentacene prior to the
film deposition. As-purchased pentacene powder was purified by
means of physical vapor deposition in a temperature gradient in
the presence of a stream of argon gas as described in
Ref.~\cite{deBoer04}. After this step, the pentacene powder was
loaded in the Knudsen cell in the UHV system and was further
purified by heating it at a temperature just below the sublimation
temperature for several days. The film thickness was determined by
calibrating the pentacene deposition rate {\it ex-situ}, using an
atomic force microscope (AFM).

Potassium doping was achieved by exposing the films to a constant
flux of K atoms generated by a current-heated getter source. The
source was calibrated and the potassium concentration determined
by means of an elemental analysis performed on PEN films doped at
several doping levels, using {\it ex-situ} Rutherford
backscattering (RBS). As shown in the top inset of
Fig.~\ref{fig:fig1} the ratio of K atoms to PEN molecules, N$_{\rm
K}$/N$_{\rm PEN}$, increases linearly with increasing the doping
time, as expected. Deviations from linearity --approximately
10-20\%-- are due to inhomogeneity of the potassium concentration.

\section{Transport properties of K$_x$PEN}

\subsection{Electronic transport through high structural-quality K$_x$PEN films}

\begin{figure}
\includegraphics[width=1\columnwidth]{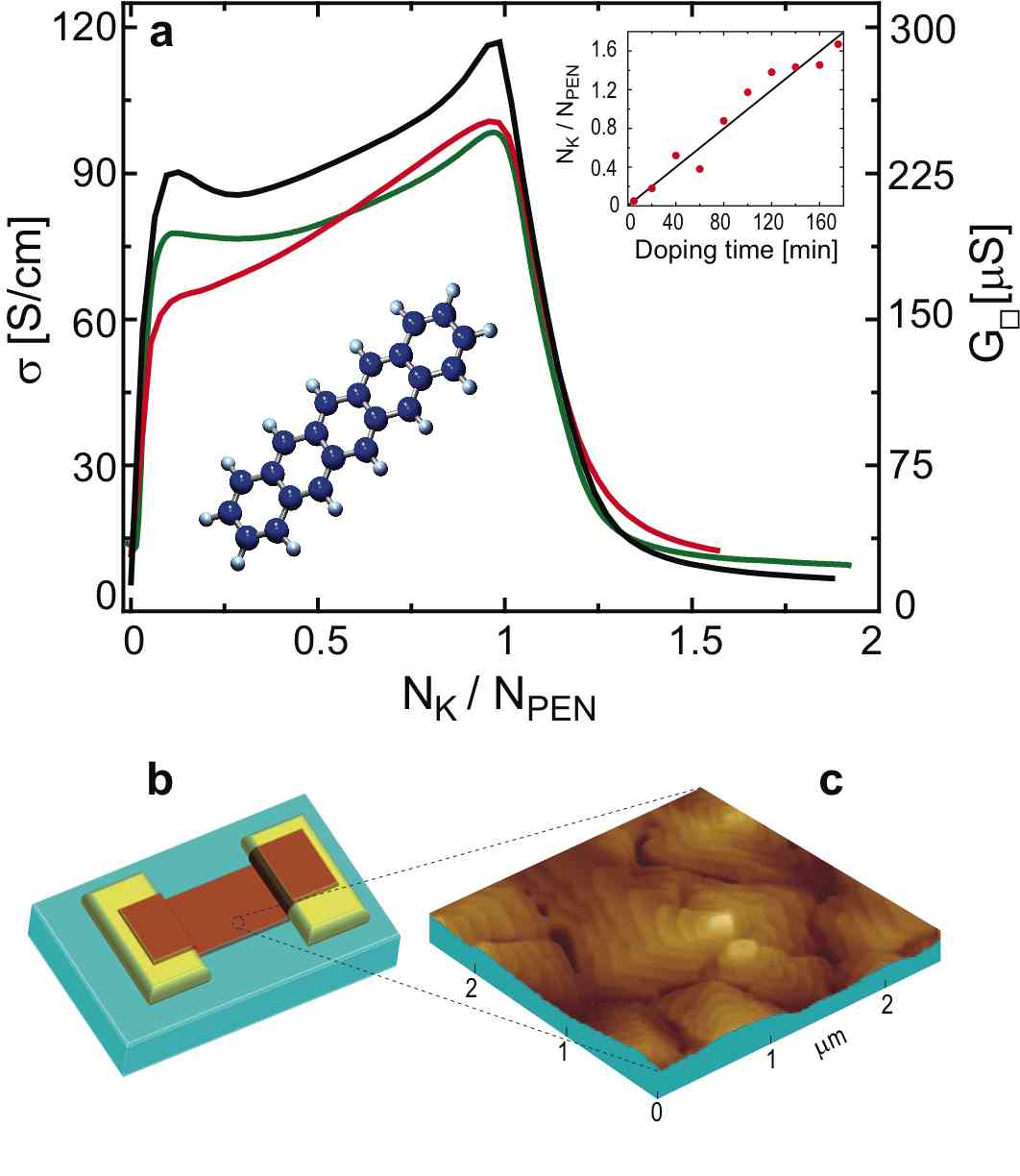}
\caption{(a) Conductivity $\sigma$ and square conductance
$G_{\Box}$, of three different K-doped PEN films, as function of
the ratio N$_{\rm K}$/N$_{\rm PEN}$; under the curves a pentacene
molecule.  Inset: N$_{\rm K}$/N$_{\rm PEN}$ as a function of
doping time. Schematic view (b) of  our set up and (c) atomic
force microscopy image of a high quality undoped PEN film showing
large crystalline grains. } \label{fig:fig1}
\end{figure}

The conductance of K$_x$PEN films is measured \textit{in situ} in
a two terminal measurement configuration with a contact separation
of approximately 175 $\mu$m (see Fig.~\ref{fig:fig1}b). The
dependence of the conductivity on the potassium concentration,
hereafter referred to as the "doping curve", is determined for
different PEN films, as a function of the ratio of K atoms to PEN
molecules. The doping curves for different samples are very
similar, as shown in Fig.~\ref{fig:fig1}a. Upon doping, the
conductivity initially increases rapidly up to a value of $\sigma
\sim 100 ~S/cm$ --in the same range as the conductivity of
metallic K$_3$C$_{60}$~\cite{Palstra92}. Upon doping further, the
conductivity continues to increase more slowly, reaches a maximum
at a concentration of one K per PEN and then drops sharply back to
the value of the undoped PEN film. All of the more than 40 films
that we have investigated exhibit a similar behavior.

The observed suppression of the conductivity of pentacene films at
high doping (for potassium concentrations higher than 1 K/PEN)
allows us to exclude the possibility that the conduction of the
intercalated films observed in the experiments is due to an
experimental artifact, for instance the formation of a potassium
layer on top of the pentacene film. In fact, at doping higher than
1 K/PEN the measured conductance, and its temperature dependence,
is essentially identical to what is measured for pristine films.

\begin{figure}
\includegraphics[width=0.85\columnwidth]{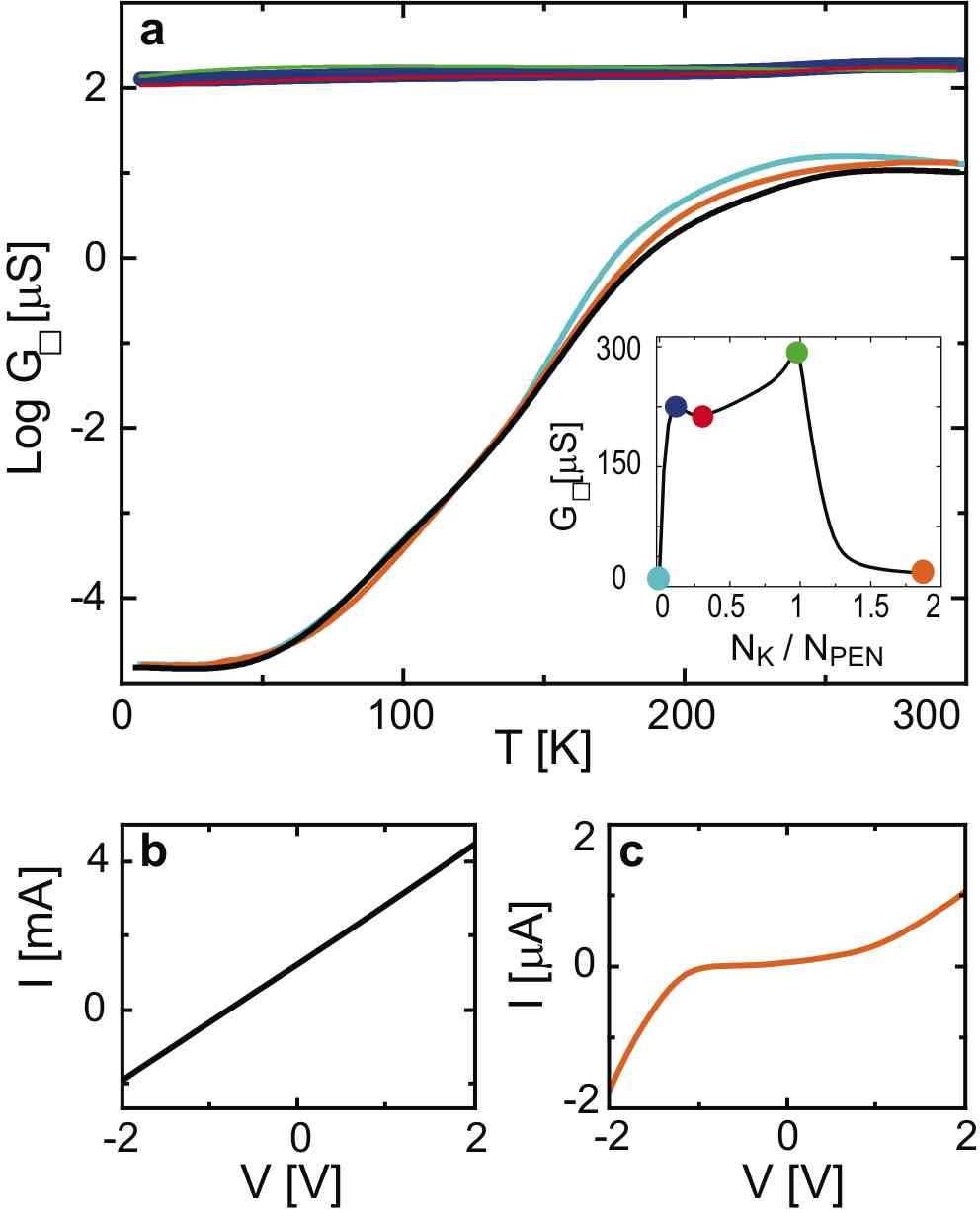}
\caption{Temperature dependence of the conductance of potassium
intercalated pentacene films. (a) The colored dots in the inset of
(a) indicate the doping level at which the temperature dependent
conductivity measurements with corresponding color were performed.
In black the temperature dependent conductivity of the Si
substrate is shown. The low temperature (5K) $I$-$V$
characteristics of K$_x$PEN in the (b) conducting and (c)
highly-doped insulating state.}
 \label{fig:Tdep}
\end{figure}

To understand the nature of conduction of pentacene films at high
carrier density we measured the temperature dependence of the
conductivity for different values of potassium concentration (see
Fig.~\ref{fig:Tdep}a). Pristine PEN films have a very low
conductivity and the measured conductance of undoped films is
dominated by transport through the substrate's 2 $\mu$m thick Si
top-layer. The measured conductivity decreases rapidly with
lowering temperature, as expected, confirming that undoped ($x$=0)
pentacene films are insulating. On the contrary, in the highly
conductive state --for $x$ between 0.1 and 1 --the conductance of
the films remains high down to the lowest temperature reached in
the experiments ($\sim$ 5 K), indicating a metallic state. When
the potassium concentration is increased beyond approximately 1
K/PEN, the conductivity again decreases rapidly with lowering
temperature, indicating a re-entrance into an insulating state.
The metallic and insulating nature of pentacene thin-films at
different potassium concentrations is confirmed by measurements of
volt-amperometric characteristics ($I$-$V$ curves) at 5 K. For $x$
between 0.1 and 1 the films exhibit linear $I$-$V$ characteristics
(Fig.~\ref{fig:Tdep}b, as expected for a metal. In the highly
doped regime (for $x>1$), on the contrary, the insulating state
manifests itself in strongly non-linear $I$-$V$ curves and
virtually no current flowing at low bias (Fig.~\ref{fig:Tdep}c).
Therefore, the data clearly show that pentacene films undergo a
metal-insulator transition as the density of potassium is
increased past one atom per molecule. Since in the overdoped
regime the conduction occurs through the Si layer of the SOI
substrate, it is not possible to gain specific information about
the properties of the insulating KPEN films --for instance, to
determine the electronic gap from measurements of the activation
energy of the conductivity-- by studying {\it dc} transport on our
samples.

\subsection{The effect of structural disorder on the transport properties
of K$_x$PEN}

The high structural quality of the films proves to be the
essential ingredient necessary to obtain K$_x$PEN films which
exhibited metallic conductivity. We find that the quality of
pentacene thin films is highly sensitive to the choice of the
substrate material and sufficient quality can be achieved by using
a hydrogen-terminated Si surface. To illustrate this important
technical point, we show here that the structural quality of
films deposited on a SiO$_{2}$ surface has a very large impact on
their electronic transport, with low quality resulting in considerably 
poorer electrical properties.

\begin{figure}
\centering
\includegraphics[width=1\columnwidth]{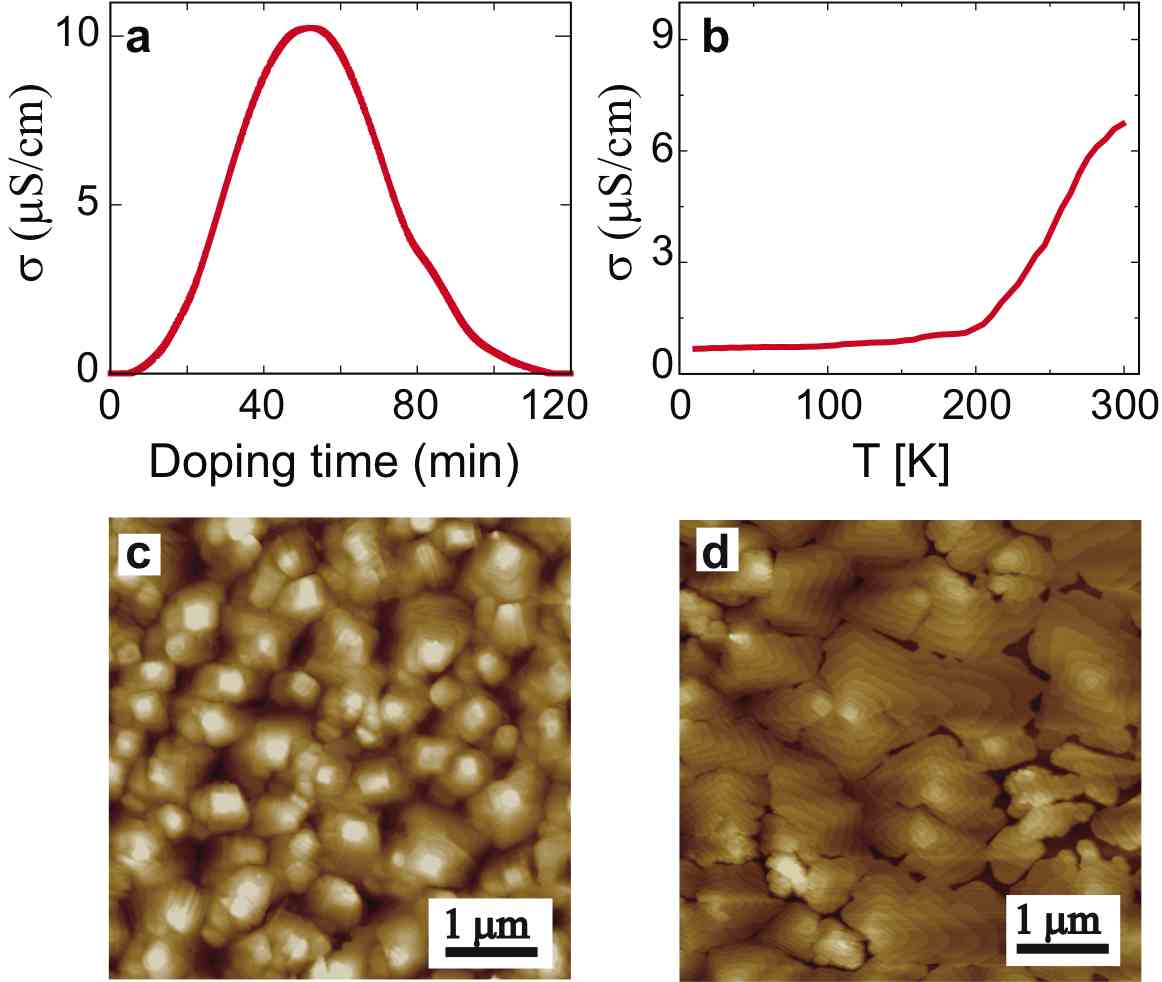}
\caption{(a) Conductivity, $\sigma$ measured at room temperature
as a function of doping time for a 25nm thick pentacene film
deposited on SiO$_{2}$. (b) Temperature dependence of the
conductivity for a pentacene film grown on SiO$_{2}$ and doped
into the highest conductivity state. The conductivity is rapidly
decreasing with lowering the temperature as it is typical for an
insulator. (c and d) AFM images of pentacene films grown on
SiO$_{2}$ and on H-terminated Si. Small and randomly oriented
grains with large height fluctuations are observed when the PEN
films are deposited on SiO$_{2}$ (c), whereas PEN films of similar
thickness deposited on Si consist of large crystalline grains with
a common relative orientation and only relatively small
fluctuations in height (d). } \label{fig:fig1sm}
\end{figure}

Fig.~\ref{fig:fig1sm}a shows the doping curve of PEN films
deposited onto 300nm SiO$_{2}$ that was thermally grown on a Si
substrate. For these films, the maximum conductivity that we
measured experimentally is several orders of magnitude lower than
the conductivity measured for films deposited on a Si surface. In
addition, (on SiO$_{2}$) the conductivity  was always observed to
decrease rapidly with lowering temperature, i.e. the potassium
intercalated films are always insulating (see
Fig.~\ref{fig:fig1sm}b). Both the magnitude and the temperature
dependence of the conductivity that we measured on SiO$_{2}$
substrates are comparable to results obtained in earlier work
reported in the literature.

\begin{figure}
\centering
\includegraphics[width=\columnwidth]{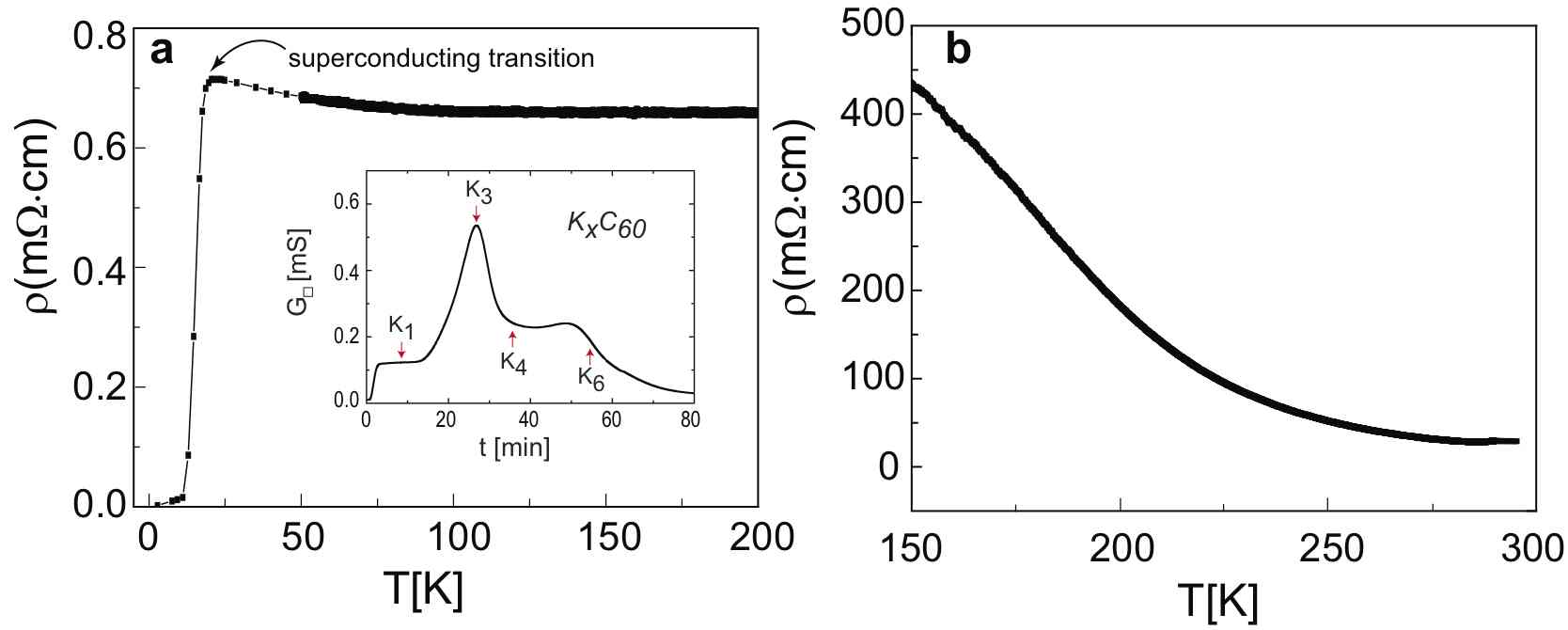}
\caption{(a) Temperature dependence of the resistivity of a
high-quality $K_{3}C_{60}$ film grown on Si. As expected, the
resistivity exhibits a superconducting transition at 18 K. The
inset shows the doping dependence of the conductivity. The
conductance peak is typical of $K_{3}C_{60}$. (b) Temperature
dependence of the resistivity of a $K_{3}C_{60}$ film grown on
SiO$_{2}$ showing insulating behavior. } \label{fig:fig2sm}
\end{figure}

We attribute the difference in the electrical behavior observed
for films deposited on Si and SiO$_{2}$ substrates to the
difference in film morphology, which we have analyzed using an
atomic force microscope. Figure~\ref{fig:fig2sm} shows AFM images
of two pentacene films of similar thickness deposited on the
hydrogen terminated Si (Fig.~\ref{fig:fig2sm}a and b) and on the
SiO$_{2}$ surface (Fig.~\ref{fig:fig2sm}c and d). It is apparent
that very different morphologies are observed for the two
substrates. PEN films deposited on Si surfaces exhibit large
crystalline grains with a common relative orientation and only
relatively small fluctuations in height. On SiO$_{2}$, on the
contrary, the grains are much smaller, randomly oriented and they
exhibit much larger height fluctuations.

This conclusion is consistent with past studies~\cite{Ruiz03} showing 
that the growth and morphology of pentacene films
is strongly influenced by the substrate surface. Specifically, for
pentacene films grown on SiO$_{2}$, a high density of nucleation
centers was observed, leading to the growth of small islands and
to a high concentration of grain boundaries. On the hydrogen
terminated silicon surface, on the other hand, the much smaller
density of nucleation centers results in significantly larger
islands and in a reduced density of grain boundaries. Note that
the critical influence of the film morphology on the electrical
characteristics of electron-doped pentacene films is also
supported by recent experiments studying the conduction of
rubidium-intercalated pentacene films deposited on
glass~\cite{Kanedo05}. In that work, as-doped films exhibited an
insulating temperature dependence of the conductivity. However, by
performing a high-temperature annealing on the doped films, which
results in an improved morphological quality, metallic behavior
was also observed.

The sensitivity of the morphology of pentacene films to the
substrate, together with the resulting effects on the electronic
properties, is common to films of many conjugated molecules. In
fact, a similar sensitivity was found in our earlier work on the
electronic properties of alkali doped Metal-Phthalocyanine (MPc)
films~\cite{Craciun06}. Specifically, for films of CuPc, NiPc,
ZnPc, FePc, and MnPc, the maximum conductivity which can be
achieved upon alkali doping when the films are deposited on
SiO$_{2}$ substrates is several orders of magnitude lower than the
conductivity measured for films deposited on a Si surface, and has
always an insulating temperature dependence. Also for alkali doped
C$_{60}$ films, it has been observed that the surface termination
of the substrate affects the morphology and the electronic
transport properties of the films. As illustrated in
Fig.~\ref{fig:fig2sm}a, the resistivity of K$_{3}$C$_{60}$ films
grown on Si shows a low resistivity at low temperature and a
transition to a superconducting state. On the contrary, the
K$_{3}$C$_{60}$ films grown on SiO$_{2}$ have significantly higher
resistivity, exhibiting thermally activated transport (see
Fig.~\ref{fig:fig2sm}b), without a superconducting transition.

\section{Interpretation in terms of a Mott state of K$_1$PEN}

The most striking aspect of our observations, namely a sharp
decrease in the conductivity starting at a carrier concentration
of one electron per molecule concomitant with the re-entrance into
an insulating state, has not been reported in earlier experiments
on intercalated pentacene (in which the density of intercalants
could not be
determined~\cite{Minakata93,Matsuo04a,Matsuo04b,Kanedo05,Fang05}),
nor in studies of pentacene field-effect transistors with gate
electrolytes (in which a metallic state has not been
observed~\cite{Panzer05}). It implies that, contrary to the case
of pentacene devices used in plastic electronics, the electronic
properties of pentacene films at high carrier density cannot be
described in terms of non-interacting electrons. In fact, even
though it is known that pentacene molecules can accept only one
electron and that doubly negatively charged pentacene ions do not
exist~\cite{Szczepanski95,Halasinski00} (i.e., in our films charge
transfer from the potassium atoms saturates at 1 K/PEN), a carrier
concentration of one electron per molecule corresponds to a
half-filled band, and, for non-interacting electrons, should
result in a metallic state. Therefore, interactions need to be
invoked in order to explain our observations.

An established scenario for the formation of an insulating state at
half-filling is the one of a Mott insulator emerging from strong 
electron-electron interactions~\cite{Imada98,Lof92}. In a Mott
insulator a strong Coulomb repulsion prevents two electrons to
occupy the same pentacene molecule. Since at half-filling the
motion of electrons necessarily requires double occupation of
molecular sites, electron transport is suppressed and the system
becomes insulating. This scenario is usually modelled
theoretically using a Mott-Hubbard Hamiltonian, which includes a
kinetic energy term described within a tight-binding scheme, and
an on-site repulsion term. The Mott state occurs when this
repulsion ($U$) is larger than the bandwidth $W$ (determined by
the tight-binding hopping amplitudes $t$). In this case the
half-filled band splits into a lower (completely filled) and an
upper (completely empty) Hubbard band, separated by a (Mott) gap
of the order of $U$, when the interactions are strong. It is
realistic that this scenario is realized in a molecular solid such
as pentacene, in which the bandwidth is expected to be small owing
to the absence of covalent bonds between the molecules.

To substantiate the Mott-insulator hypothesis we have analyzed the
electronic structure using density functional theory (DFT)
calculations to extract the parameters of the Mott-Hubbard model.
A main difficulty in doing this is that the structural knowledge
of the intercalated films is incomplete, as our ultra-high vacuum
set-up is not equipped to perform {\it in situ} structural
characterization, and {\it ex situ} characterization is impeded by
oxidation of potassium when the sample is extracted from the
vacuum system where the films are prepared. Therefore, for the DFT
calculations we take advantage of the existing structural
information on intercalated pentacene compounds and we determine
the stable crystal structure of K$_1$PEN using a computational
relaxation procedure which refines the positions of all the atoms
in the unit cell.

\subsection{Structural Details of K$_1$PEN}

It is well known from previous structural studies on pentacene
films that the herringbone arrangement of the molecules is
preserved when pentacene is intercalated with
iodine~\cite{Minakata91,Minakata92,Ito04} or with different alkali
atoms~\cite{Minakata93,Matsuo04a,Matsuo04b,Kanedo05,Fang05}, and
that intercalation takes place between the pentacene layers. It is
also known that intercalation is accompanied by a considerable
expansion of the unit cell $c$-axis (by an amount close to the
radii of the intercalated ions) while the in-plane lattice
parameters $a$ and $b$ are only minorly
affected~\cite{Matsuo04a,Matsuo04b}.

A reliable estimate of the length of the expanded $c$-axis is
given by the sum of the radius of the alkali ion and the pristine
$c$-axis parameter: the $c$-axis lattice constants that are
obtained in this way for, e.g., RbPEN and CsPEN are within 2\% of
the experimental values. For K$_1$PEN we construct the lattice
parameters starting from two different polymorphs  (one with
$c=14.33$ \AA and the other with $14.53$ \AA), using a K$^+$ ionic
radius of 1.33 \AA. The structures of the two polymorphs are taken
from the experimental results of Ref. \cite{Mattheus01}. They
differ slightly in the packing of the pentacene molecules, which
enables us to study the influence of realistic variations of the
packing on the electronic structure. The actual size of the c-axis
in the K$_1$PEN is not critical. Our relaxation and band-structure
computations were checked for values up to $8 \%$ larger and
smaller than the estimated $c$-axis parameters. We found that even
such relatively large variations of $c$ do not affect our main
results (i.e., the values of the calculated bandwidth $W$ and on
site repulsion $U$), because the interaction between adjacent
pentacene layers is weak.

After constructing the unit cell of potassium-intercalated
pentacene, using the information above to fix $a$, $b$ and $c$, we
refine the positions of the atoms by a computational relaxation
procedure. For all the electronic structure calculations we used
the Vienna Ab Initio simulation package
(VASP)~\cite{Kresse93,Kresse96} with projector augmented waves
(PAW)~\cite{Kresse99} and the PW91 density
functional~\cite{Perdew92}. The self-consistent calculations were
carried out with an integration of the Brillouin using the
Monckhorst-Pack scheme with a 6x6x4 $k$-points grid and a smearing
parameter of 0.01 eV, and a plane-waves basis set with a cutoff
energy of 550 eV. To determine the stable structure of K$_1$PEN
all the atoms positions in the unit cell are relaxed using a
conjugate-gradient method. To avoid possible energy barriers we
used a number of different initial configurations. In the
relaxation procedure first the forces on the K-ions are calculated
and then the K positions are relaxed. We observe that the dopants
move into high symmetry positions in the plane between the
pentacene layers. In the next step the positions of $\it all$
atoms in the unit cell are relaxed --including the ones of the two
PEN molecules. The final stable structure is the same for all
different initial configurations. \footnote{The data file with the resulting
structure for both polymorphs is available directly from the
authors.}

\begin{figure}
\includegraphics[width=1\columnwidth]{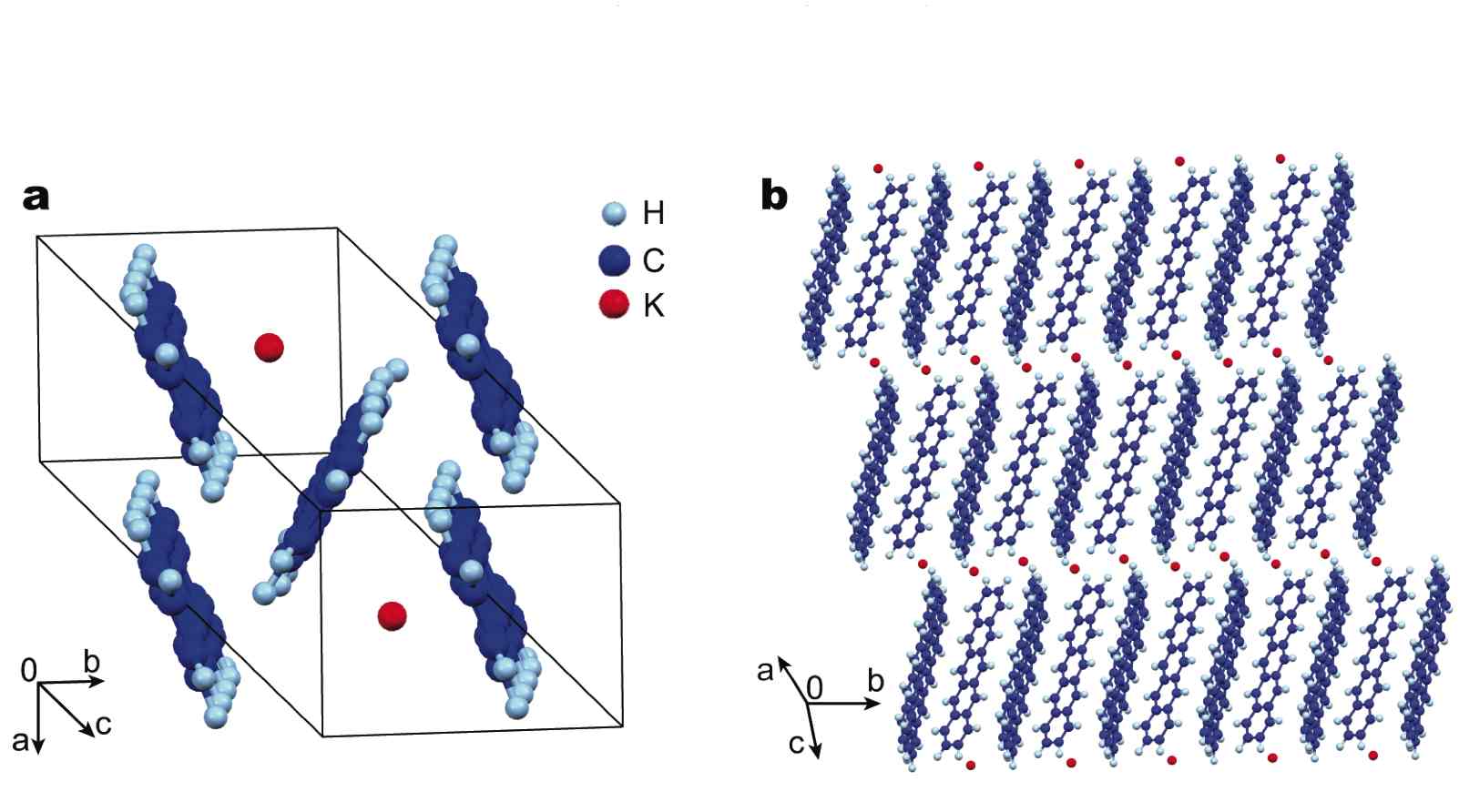}
\caption{Crystal structure of potassium-intercalated pentacene
K$_1$PEN obtained from {\it ab initio} computational relaxation
with (a) showing the herringbone of the PEN molecules and K atoms
in the unit cell and (b) is a side view of the stacked layers of
PEN and K, illustrating the potassium intercalation in between the
molecular planes. The unit cell parameters are $a,b,c=6.239,
7.636, 15.682$ \AA, $\alpha, \beta, \gamma=76.98, 88.14,
84.42^{\circ}$} \label{fig:fig3}
\end{figure}

The optimized structure of K$_1$PEN is shown in
Fig.~\ref{fig:fig3}. We checked the reliability of the relaxation
procedure on undoped pentacene and found that the calculated
structure indeed corresponds to the actual, known crystal
structure of the material. In K$_1$PEN there are two inequivalent
PEN molecules per unit cell, just as in the undoped compound.
Intercalation changes the detailed molecular orientations in the
unit cell, see Fig.~\ref{fig:fig3}, but we do not observe the
formation of superstructures such as for instance molecular
dimers. For the two distinct pentacene
polymorphs~\cite{Mattheus01} for which we have performed the
relaxation procedure, we found that the conclusions on electronic
bandwidths and Coulomb interactions that will be presented
hereafter hold equally well.

\subsection{Electronic Structure and Electronic Correlations in K$_1$PEN}

\begin{figure}
\includegraphics[width=1\columnwidth]{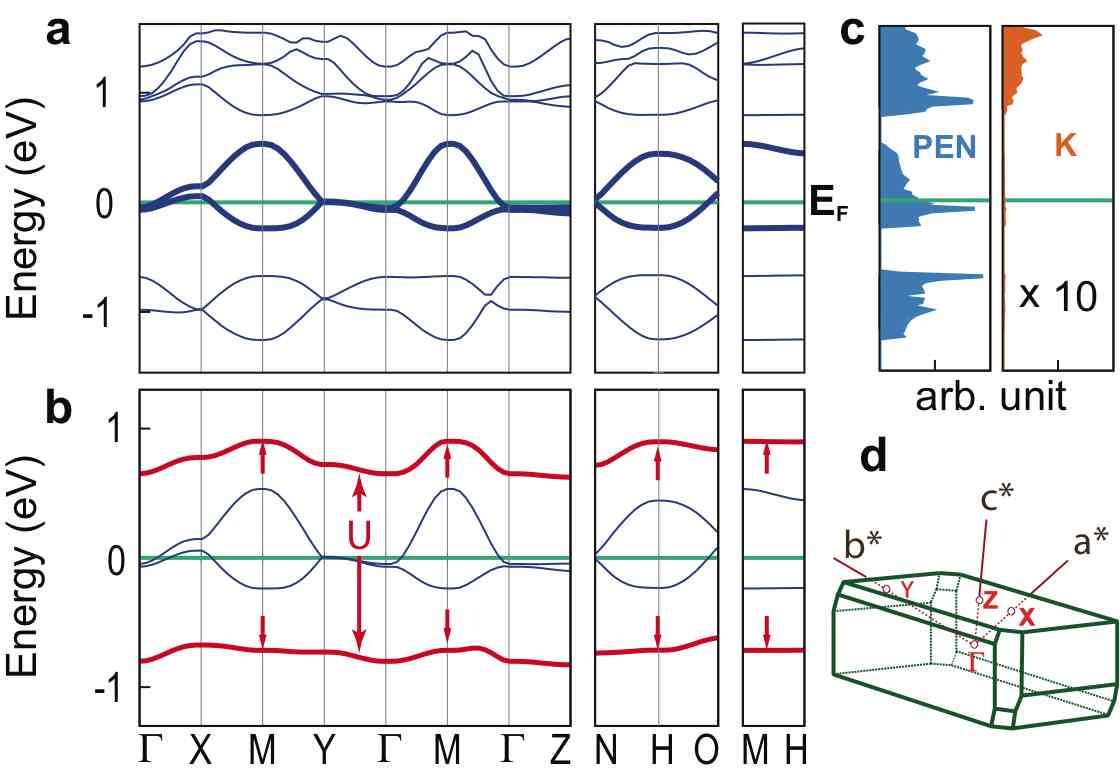}
\caption{Results of electronic structure calculations for
K$_1$PEN. (a) Single-particle band-structure~\cite{Remark2}, with
the Fermi-level $E_{F}$ (green line) as the zero of energy.
Valence and conduction band are indicated by the thick blue lines.
(b) Carbon (blue) and potassium (orange) projected density of
states. (c) Tight-binding fit to the valence and conduction band
(blue thin lines) and the resulting lower and upper Hubbard band
from a mean field analysis of the corresponding Hubbard
Hamiltonian with $U=1.45$ eV (red thick lines). The arrows
indicate the opening of the Hubbard gap. (d) Reciprocal lattice
vectors and the first Brillouin zone of K$_1$PEN.}
 \label{fig:fig4}
\end{figure}

Figure~\ref{fig:fig4}a shows the DFT band structure of K$_1$PEN
together with the projected density of states on the pentacene and
potassium orbitals for the polymorph associated with the relaxed
structure of Fig.~\ref{fig:fig3}. The Fermi energy lies in the
middle of a half-filled band that is entirely of pentacene
character, originating from its LUMO. The potassium derived
electronic states are present only at much higher energy,
demonstrating that little hybridization takes place and that the
role of the potassium atoms is limited to transferring its
electrons to the pentacene molecules. The total bandwidth is
$W$=0.7 eV. From a tight-binding fit of the band dispersion (see
Fig.~\ref{fig:fig4}b) we extract the hopping amplitudes $t_{ij}$
that enter the kinetic energy part of the Mott-Hubbard Hamiltonian

\begin{eqnarray}
H = \sum_i e_i n_i + \sum_{\langle ij \rangle, \sigma}  t_{ij} (c_{i, \sigma}^{\dagger}c_{j, \sigma}^{} + h.c.) +U \sum_i n_{i,\uparrow} n_{i,\downarrow}, \nonumber
\end{eqnarray}
where we have two molecules in the unit cell, with on-site energy
$e_i$, the electron creation (annihilation) operators on site $i$
are $c_{i,\sigma}^{\dagger}$ ($c_{i,\sigma}^{}$), with $\sigma$
the electron spin,  $h.c.$ the hermitian conjugate,
$n_{i,\sigma}=c_{i,\sigma}^{\dagger}c_{i,\sigma}$,
$n_i=\sum_\sigma n_{i,\sigma}$ and $U$ the effective Coulomb
interaction between two electrons on the same molecule. The
hopping integrals $t_{ij}$ are different in different directions
and between nearest and next nearest neighbor molecules, see
Table~\ref{table:tight}. The resulting electronic band-structure
(Fig.~\ref{fig:fig4}a)  displays only very minor differences for
the two stable polymorphs and within our accuracy the
tight-binding parameters are the same.

\begin{table}
\begin{tabular}{|l|r||l|r||l|r|}
\hline
parameter &   meV  & parameter &   meV& parameter &   meV \\
\hline
 $e$           &  39 & & & & \\
 \hline
 $t_{a}$           & -33 & $t_{b}$           & -11 & $t_{c}$           &  1 \\
 $t_{2a}$          & -1 & $t_{a+b}$         &  1 & $t_{a-b}$         & -9\\
 $t_{a+c}$         & -6 & $t_{b+c}$         &  3 & $t_{a+b+c}$       & -5 \\
 $t_{(a+b)/2}$     & -96 & $t_{(a-b)/2}$     &  90 & $t_{(3a+b)/2}$    & -4 \\
 $t_{(3a-b)/2}$    &  9 & $t_{3a/2+b/2+c}$  &  1 & $t_{a/2+3b/2+c}$  & -3 \\
 $t_{3a/2+3b/2+c}$ & -2 & $t_{3(a+b)/2}$    & -3 & $t_{3(a-b)/2}$    &  2 \\
 \hline
\end{tabular}
\caption{Tight-binding fit parameters to the {\it ab initio}
band-structure of the half-filled conduction/valence band of
$KPEN$. The on-site energy difference is denoted by $e$ and the
hopping parameters along the $a$, $b$ and $c$ axes are denoted by
$t$.}
 \label{table:tight}
\end{table}

In order to determine the relative strength of electronic
correlations and to compute the magnetic exchange interactions,
the on-site Coulomb interaction $U_{bare}$ for two electrons on
the same pentacene molecule is determined, using the techniques
described in Ref.~\cite{Brocks04}. For this the total energy of
neutral and charged pentacene molecules is calculated by density
functional calculations in the local density approximation using
GAMESS with a DZVP basis-set~\cite{Schmidt93}. The bare value of
the Coulomb interaction is found to be $U_{bare}=3.50$ eV. In the
solid this value is screened, leading to a lower value
$U$~\cite{Giovannetti08,Brink95,Meinders95}. From the eigenvalues
of the charged molecule that is placed inside a cavity of an
homogeneous dielectric medium with dielectric constant of 3.3
using the SS(V)PE model~\cite{Brocks04,Chipman00}, one finds
$U=1.45$ eV. We have also performed an independent estimate for
the value of $U$ by considering the difference between the band
gap of pristine pentacene from density functional calculations
(0.7 eV) and its  experimental value (2.2 eV~\cite{Silinsh82}),
which gives $U \approx 1.5$. These two values, determined in two
very different ways, are remarkably close. Very similar values for
$U$ are found also for the second polymorph used in our
calculations, indicating that these values are not very sensitive
to differences in the structure~\cite{Remark}.

From a straightforward self-consistent mean-field decoupling
computation on the resulting Hubbard Hamiltonian the ground state is found
to be a N\'eel ordered antiferromagnet, with a charge gap of 1.23
eV. The antiferromagnetic exchange between neighboring molecules
in the plane is $J=4 t_{(a \pm b)/2}^2/U\simeq 290 K$. This value
is actually an underestimation of the Heisenberg exchange, as
certainly a nearest neighbor Coulomb interaction $V$ is also
present, which has the effect of increasing the value of exchange
by a factor to $U/(U-V)$~\cite{Eder95}. We find that the coupling between
molecules in neighboring planes, $J_{\perp}$, is four orders of
magnitude lower than the in-plane $J$. Consequently K$_1$PEN is a
quasi two-dimensional antiferromagnet. Finally, an
antiferromagnetic exchange of $\simeq 40 K$ is also present
between in-plane next nearest neighbor molecules along the
$a$-axis, leading to a weak frustration of the magnetic N\'eel
ordering.

An alternative procedure to take the electronic correlations into
account, used in band structure calculations of strongly
correlated transition metal oxides, is to incorporate the {\it
atomic} Hubbard $U$ directly in a LDA+U scheme. For molecular
crystals, however, this approach would require a novel
computational scheme that uses a basis set of orthonormal
molecular orbitals so that a {\it molecular} $U$ can be
incorporated.

\section{Discussion and conclusions}

Using the results of the electronic structure calculations we are
now in a position to validate the Mott-state hypothesis. With a
ratio $U/W \simeq 2.1$, electron-electron interactions cause the
splitting of the LUMO band and the opening of a Mott gap, as shown
in Fig.~\ref{fig:fig4}c. The gap explains the observed re-entrance
to the insulating state. Note that the Mott-state scenario also
explains why the insulating state is only observed for a potassium
concentration of 1.1-1.2 atoms per molecule (and not at exactly
one; see Fig.~\ref{fig:fig1}a). In fact, at exactly one potassium
per pentacene molecule, non-uniformity in the {\it potassium}
concentration --estimated to be approximately 10-20\% in our
films-- effectively dopes the Mott insulator causing the
conductivity to remain large. However, even in the presence of
non-uniformity, a potassium concentration slightly larger than 1
K/PEN results in a uniform {\it electron} concentration exactly
equal to one electron per molecule, since, as we mentioned
earlier, only one electron (and not two) can be donated to each
pentacene molecule~\cite{Szczepanski95,Halasinski00}. It should be
noted that imperfections in the material, either due to disorder
of the dopants or in the molecular arrangements will lead to the
presence of both disorder in the bandwidth and the presence of a
disorder potential. In general the physics of disordered
Mott-Hubbard systems is very rich~\cite{Belitz94}, but it is not a
priori clear how relevant disorder will be in the present
situation as we find that the Mott state in K$_1$PEN is stabilized
by a substantial electronic gap.

Since the coupling between pentacene molecules in different layers
is very small and the electron-electron interaction is
sufficiently large, the low-energy effective electronic
Hamiltonian of the K$_x$PEN reduces to the well-known
two-dimensional $tJ$-model~\cite{Imada98} with $t/J \approx 3-4$.
Interestingly, the same $tJ$-model in the same coupling regime
describes another important class of materials, namely strongly
correlated cuprate superconductors such as
La$_{2-x}$Sr$_x$CuO$_4$. An apparent difference between these
classes of materials is that it is generally excepted that in
doped organics the formation of lattice-polarons plays a very
important role.

We conclude that temperature-dependent transport measurements and
theoretical calculations consistently indicate that at a doping
concentration of one potassium ion per molecule
potassium-intercalated pentacene is a strongly correlated Mott
insulator, whose electronic properties are dominated by
electron-electron interactions. An immediate consequence is the
emergence of magnetism. Our calculations show that the magnetic
interactions are dominated by a large, positive magnetic exchange
$J=4t^2/U \simeq$ 290 K between electrons on nearest neighbor
molecules in the same pentacene layer. We predict that K$_1$PEN is
therefore an antiferromagnet. In fact, experimental indications
for the presence of antiferromagnetism in intercalated pentacene
have been reported in magnetic susceptibility measurements
performed in the past~\cite{Mori97}, albeit at very low
temperature.\\

\section{Acknowledgements}

This work was supported by the Foundation for Fundamental Research
on Matter (FOM), the Royal Dutch Academy of Sciences, the NWO
Vernieuwingsimpuls, NanoNed and the Stichting Nationale
Computerfaciliteiten. We are grateful to the FOM Institute for
Atomic and Molecular Physics (AMOLF) for the RBS analysis of our
samples.

%%%%%%%%%%%%%%%%%%%%%%%%%%%%%%%%%%%%%%%%%%%%%%%%%%%%%%%%%%%%%%%%


\begin{thebibliography}{99}
\bibitem{Remark2} The points in the Brillouin zone are $\Gamma=(0,0,0), X =(\frac{1}{2},0,0), M=(\frac{1}{2},\frac{1}{2},0), Y=(0,\frac{1}{2},0), Z=(0,0,\frac{1}{2}), N=(0,\frac{1}{2},\frac{1}{2}), H=(\frac{1}{2},\frac{1}{2},\frac{1}{2})$ and $O=(\frac{1}{2},0,\frac{1}{2})$.
\bibitem{Remark} In an alternative crystal structure for intercalated pentacene, see Ref.~\cite{Hansson06}, a still smaller value of the bandwidth is found, implying an even larger correlation gap.
\end{thebibliography}
\end{document}